\documentclass[journal,transmag]{IEEEtran}
\ifCLASSINFOpdf
   \usepackage[pdftex]{graphicx}
\else
		\usepackage{graphicx}
\fi
\usepackage{amsmath}
\begin{document}
%
\title{Low Power Microwave Signal Detection With a Spin-Torque Nano-Oscillator in the Active Self-Oscillating Regime}


\author{\IEEEauthorblockN{Steven Louis\IEEEauthorrefmark{1},~\IEEEmembership{Student Member,~IEEE},
Vasyl Tyberkevych\IEEEauthorrefmark{2}, ~\IEEEmembership{Member,~IEEE},
Jia Li\IEEEauthorrefmark{1}, ~\IEEEmembership{Senior Member,~IEEE}\\
Ivan Lisenkov\IEEEauthorrefmark{3,4},
Roman Khymyn\IEEEauthorrefmark{5},
Elena Bankowski\IEEEauthorrefmark{6},
Thomas Meitzler\IEEEauthorrefmark{6},
Ilya Krivorotov\IEEEauthorrefmark{7},\\ and
Andrei Slavin\IEEEauthorrefmark{2},~\IEEEmembership{Fellow,~IEEE}}%
\IEEEauthorblockA{\IEEEauthorrefmark{1}Department of Electrical and Computer Engineering, Oakland University, Rochester, MI 48309, USA}
\IEEEauthorblockA{\IEEEauthorrefmark{2}Department of Physics, Oakland University, Rochester, MI 48309, USA}
\IEEEauthorblockA{\IEEEauthorrefmark{3}Kotelnikov Institute of Radio-engineering and Electronics of RAS, Moscow 125009, Russia}
\IEEEauthorblockA{\IEEEauthorrefmark{4}Department of Electrical Engineering and Computer Science, Oregon State University, Corvallis, OR 97331, USA}
\IEEEauthorblockA{\IEEEauthorrefmark{5}Department of Physics, University of Gothenburg, S-405 30 Gothenburg, Sweden}
\IEEEauthorblockA{\IEEEauthorrefmark{6}U.S. Army TARDEC, Warren, MI 48397, USA}
\IEEEauthorblockA{\IEEEauthorrefmark{7}Department of Physics, University of California Irvine, Irvine, CA 92697, USA}%
\thanks{Manuscript received Month XX, 2017; revised Month XX, 2017. 
Corresponding author: Steven Louis (email: slouis@oakland.edu).}}%

\markboth{BC-05 - Intermag 2017 - International Magnetics Conference Dublin}%
{Shell \MakeLowercase{\textit{et al.}}: Bare Demo of IEEEtran.cls for IEEE Transactions on Magnetics Journals}
%



\IEEEtitleabstractindextext{%
\begin{abstract}
A spin-torque nano-oscillator (STNO) driven by a ramped bias current can perform spectrum analysis quickly over a wide frequency bandwidth. The STNO spectrum analyzer operates by injection locking to external microwave signals and produces an output DC voltage $V_{\rm dc}$ that temporally encodes the input spectrum. We found, via numerical analysis with a macrospin approximation, that an STNO is able to scan a $10~\rm GHz$ bandwidth in less than $100~\rm ns$ (scanning rate $R$ exceeds $100~\rm MHz/ns$). In contrast to conventional quadratic microwave detectors, the output voltage of the STNO analyzer is proportional to the amplitude of the input microwave signal $I_{\rm rf}$ with sensitivity $S = dV_{\rm dc}/dI_{\rm rf} \approx 750~\rm mV/mA$. The minimum detectable signal of the analyzer depends on the scanning rate $R$ and, at low $R \approx 1~\rm MHz/ns$, is about $1~\rm pW$.
\end{abstract}

}

\maketitle

\IEEEdisplaynontitleabstractindextext

%
\IEEEpeerreviewmaketitle

\section{Introduction}
\IEEEPARstart{N}{ano-sized} magnetic tunnel junctions (MTJs) have served as hard drive read heads for more than a decade. 
When a DC electric current passes through an MTJ, the exchange interaction between the conduction and localized electrons creates an additional magnetic torque, called the spin transfer torque (STT) \cite{slonczewski1996current, berger1996emission}, which may lead to self-sustained excitation of the magnetization precession in one (``free'') magnetic layer of the MTJ \cite{rippard2003quantitative, kiselev2003microwave, slavin2009nonlinear}. Such magnetic oscillators are called spin-torque nano-oscillators (STNOs) and have a number of unique properties. For example, the frequency of the generated microwave signal is easily tuned by either the bias magnetic field or the bias current amplitude. STNOs can have a tunability bandwidth as high as 10~GHz, a speed of frequency tuning as fast as 5~GHz/ns, and a maximum frequency that can exceed 65~GHz \cite{bonetti2009spin,louis2016}. Many applications related to MTJs exhibiting STT have been proposed, including GHz frequency signal generators \cite{bankowski2015magnonic}, signal modulation devices \cite{quinsat2014modulation}, microwave signal detectors (diodes) \cite{tulapurkar2005spin, prokopenko2013spin, fang2016giant}, computer memory applications\cite{everspin}, energy harvesters \cite{prokopenko2013spin, finocchio2015skyrmion, hemour2012spintronics}, logic devices \cite{khitun2010magnonic, locatelli2014spin}, spin wave generators \cite{slavin2005spin}, and others. This paper studies the viability of a novel application of STT-driven MTJs: a fast STNO-based spectrum analyzer that operates over a wide bandwidth ($10~\rm GHz$) with maximum scanning rate exceeding $100~\rm MHz/ns$ and minimum detectable signal (MDS) in the $\rm pW$ range.

\begin{figure}[ht]
\centering
\includegraphics{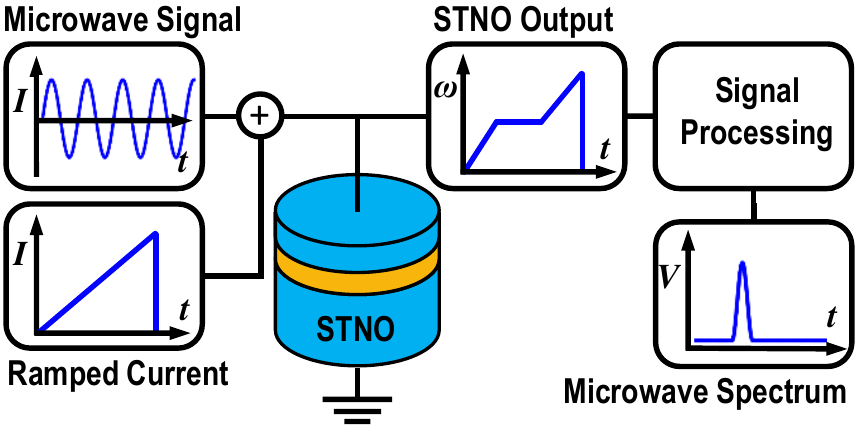}
\caption{Schematic of STNO spectrum analyzer. The ramped bias current and microwave signal to be analyzed (on the left) are applied to the STNO (in the center). The STNO output (DC component of the STNO voltage) is digitally processed to produce the time-encoded microwave spectrum of the input signal (on the right).}
\label{schematic}
\end{figure}

A schematic of the proposed STNO spectrum analyzer is shown in Fig.~\ref{schematic}. The STNO has two inputs: a ramped DC bias current, and an external microwave signal to be analyzed. The DC bias current, $I_{\rm dc}(t)$, initiates STNO microwave signal generation and continuously tunes the STNO generation frequency through the microwave detection range. When the STNO frequency matches that of the external signal, the STNO will injection lock to the external signal and produce a DC spike by the spin torque diode effect \cite{tulapurkar2005spin}. The STNO output voltage undergoes signal processing, and the temporal position of the DC spike indicates the frequency of the external signal. Thus, the frequency spectrum of the input signal is encoded in the temporal profile of the output DC voltage.

Current analytical theory does not fully describe STNO phase locking behavior while scanning a finite frequency range. Therefore, we used numerical simulations in a macrospin approximation to develop an understanding of STNO response to an external signal in this regime. By using the numerical simulations, we demonstrate that the proposed STNO spectrum analyzer performs faithful temporal encoding of the input signal spectrum and can operate at the frequency scanning rates exceeding $100~\rm MHz/ns$. The amplitude of the output DC peak increases linearly with the input microwave signal amplitude and sensitivity $S = dV_{\rm dc}/dI_{\rm rf} \approx 750~\rm mV/mA$. This property of the STNO detector based on the phase locking effect distinguishes it from conventional and spin-torque diode detectors, in which the output signal is proportional to the  input signal power (i.e., square of the microwave current). We also show that a finite rate $R$ of STNO frequency ramp leads to the appearance of the threshold microwave current $I_0(R)$, below which the output DC voltage vanishes because the ramped STNO does not have enough time to phase lock to the weak signal. The threshold current $I_0(R)$ is the main factor limiting the minimum detectable signal (MDS) of the STNO spectrum analyzer at high scanning rates $R$. For relatively slow scanning rates $R \approx 1~\rm MHz/ns$ the MDS is limited, mostly, by the thermal noise and is of the order of $1~\rm pW$. 

\section{Principle of Operation of STNO Spectrum Analyzer}

The proposed STNO spectrum analyzer employs several well-known physical effects. First, the current-induced magnetization precession in the MTJ results in oscillating dependence of the MTJ electrical resistance \cite{kiselev2003microwave},
\begin{equation}
	r_{\rm stno}(t) = R_0 - \Delta R_{\rm stno}\cos(2 \pi f_{\rm stno}t+\psi)
\label{rstno}
\,,\end{equation}
where $R_0$ is the averaged MTJ resistance, oscillation amplitude $\Delta R_{\rm stno}$ depends on the tunneling magnetoresistance (TMR) of the MTJ and amplitude of precession, $f_{\rm stno}$ is the STNO generation frequency, and $\psi$ is the oscillation phase. The generation frequency $f_{\rm stno}$ is determined by the bias current $I_{\rm dc}$ and, if the current is ramped, can be continuously tuned in a wide range \cite{rippard2003quantitative, louis2016}. In the following, we shall denote the rate of change of the STNO frequency (the scanning rate) as $R = df_{\rm stno}/dt$.

If an external microwave current
\begin{equation}
	i_{\rm rf}(t) = I_{\rm rf}\cos(2\pi f_{\rm ext} t)
\label{irfIn}
\end{equation}
is injected into the STNO, the STNO oscillations may phase-lock to this current \cite{rippard2005injection}. In the phase-locking regime, the STNO generates at exactly the external frequency $f_{\rm ext}$, while the STNO phase shift $\psi$ is determined by the internal properties of STNO and frequency mismatch between the free-running STNO frequency and the signal frequency \cite{zhou2007intrinsic, zhou2008tunable, slavin2009nonlinear}.

Mixing of the microwave current Eq.~(\ref{irfIn}) with coherent resistance oscillations Eq.~(\ref{rstno}) results in the generation of an additional DC voltage at the STNO:
\begin{equation}
	V_{\rm dc} = \langle i_{\rm rf}(t)~r_{\rm stno}(t) \rangle =\bigg(\hspace{-1.5 mm}-\frac12 \Delta R_{\rm stno} \cos(\psi) \bigg) I_{\rm rf}
\label{vdc}
\ .\end{equation}
Note, that, in contrast with the usual ``passive'' spin-torque diode effect \cite{tulapurkar2005spin}, the amplitude of the resistance oscillations $\Delta R_{\rm stno}$ in the self-oscillating regime is determined mostly by the bias STNO current $I_{\rm dc}$ and is practically independent of a weak microwave signal $I_{\rm rf}$. Therefore, the output DC voltage Eq.~(\ref{vdc}) is proportional to the {\em amplitude} $I_{\rm rf}$ (rather than the power $I_{\rm rf}^2$) of the input signal. This property distinguishes the active STNO detector from conventional quadratic diode detectors and suggests that the STNO detector may have increased sensitivity to weak signals and low minimum detectable signal (MDS) levels.

The output DC voltage Eq.~(\ref{vdc}) is generated only when the STNO is phase-locked to the external signal. If the free-running STNO frequency is sufficiently far from the microwave signal frequency, the STNO and external signal oscillations are uncorrelated, and the output voltage vanishes. In the proposed device driven by a ramped bias current $I_{\rm dc}(t)$ the free-running STNO frequency continuously varies in time and the output DC voltage appears only at a moment of time when $f_{\rm stno} = f_{\rm ext}$. Thus, the STNO spectrum analyzer temporally encodes the spectrum of an input microwave signal in the form of DC voltage peaks.

\section{Methods}

In order to test the validity of the above presented STNO spectrum analyzer principle of operation, we performed numerical simulations of the STNO free layer magnetization using the Landau-Lifshitz-Gilbert-Slonczewski equation in a macrospin approximation \cite{slavin2009nonlinear}:
\begin{equation}
	\frac{\mathrm{d}\mathbf{m}}{\mathrm{d}t} = \gamma\mathbf{m}\times \mathbf{B}_{\rm eff}+\alpha_{\rm G}\mathbf{m}\times\frac{\mathrm{d}\mathbf{m}}{\mathrm{d}t}+|\gamma|\alpha_{\rm J}I(t)\mathbf{m}\times[\mathbf{m}\times\mathbf{p}]\label{llgs}.\end{equation}
In this equation, $\mathbf{m}$ is the normalized unit-length magnetization vector, 
$\gamma=-1.76 \times 10^{11}~{\rm Hz/T}$ is the gyromagnatic ratio, $\mathbf{B}_{\rm eff}=\mathbf{B}_{\rm ext}-\mu_0 M_{\rm s} (\mathbf{m}\cdot\mathbf{\hat{z}})\mathbf{\hat{z}}$ is the effective field, $\mathbf{B}_{\rm ext}=1.5~{\rm T}$ is the external field applied in the $\mathbf{\hat{z}}$ direction, which is normal to the free layer plane, and $\mu_0 M_s=0.8~{\rm T}$ is the free layer saturation magnetization. The Gilbert damping constant is $\alpha_{\rm G}=0.01$, and $\alpha_{\rm J}=\hbar \eta_0 / (2 \mu_0M_{\rm s}eV)$, where $\hbar$ is the reduced Planck constant, the spin polarization efficiency is $\eta_0=0.35$, the input current is $I(t)$, $\mu_0$ is free space permeability, $e$ is the fundamental electric charge, and $V=3\times 10^4~{\rm nm}^3$ is the volume of the free layer (this is equivalent to a $4~{\rm nm}$ thick permalloy disk with a $50~{\rm nm}$ radius). The direction of spin current polarization was chosen as $\mathbf{p}=\cos(\beta)\mathbf{\hat{x}}+\sin(\beta)\mathbf{\hat{z}}$ with $\beta=30^\circ$. With this configuration, the STNO threshold current of microwave signal generation was was $2.32~{\rm mA}$.

The tunneling magnetoresistance of the STNO was simulated as $R(\theta)=R_0 -\Delta R_0\cos \theta$, where $R_0=1.5$~k$\Omega$ is the average resistance of the STNO, and $\Delta R_0 \cos \theta= \Delta R_0 (\mathbf{m}\cdot\mathbf{p})$ is the raw output voltage of the STNO. In this study we assumed $\Delta R_0=1$~k$\Omega$. 

The signal processing, shown on the right side of Fig.~\ref{schematic}, consists of 3 steps. In the first step, the raw output voltage of the STNO generating in the free-running regime (no input microwave current) is subtracted from the output voltage generated by a STNO in the presence of an external signal. Then, a low pass filter with cutoff frequency $\approx$15~GHz is applied. This filter removes the relatively powerful signals produced by the STNO in the 25 to 35~GHz frequency range without distorting low frequency signals. Finally, a low pass filter with a MHz range cutoff frequency of $\Delta f$ is applied. As the characteristics of the output peak produced by the STNO detector changes with the scanning rate $R$, the video bandwidth (VBW) required for output also has to be adjusted. The empirically found optimal VBW follows the rule $\Delta f = \tau R$, where $\tau \approx 2.6$~ns for the chosen STNO parameters. The two stage filter configuration was chosen to limit filter distortion while maintaining computational efficiency. In an experimental setup, the output voltage $V_{\rm dc}$ after filtering will have a low frequency and, thus, can be processed further in the digital domain.

Note that to induce an STNO generation frequency ramp in our simulations, we held the bias magnetic field constant while ramping the bias current. An alternative configuration would be to hold the bias current constant while linearly increasing the magnetic field. We chose the first configuration as it is more easily realized experimentally.

\section{Results}

\begin{figure}[ht]
\centering
\includegraphics{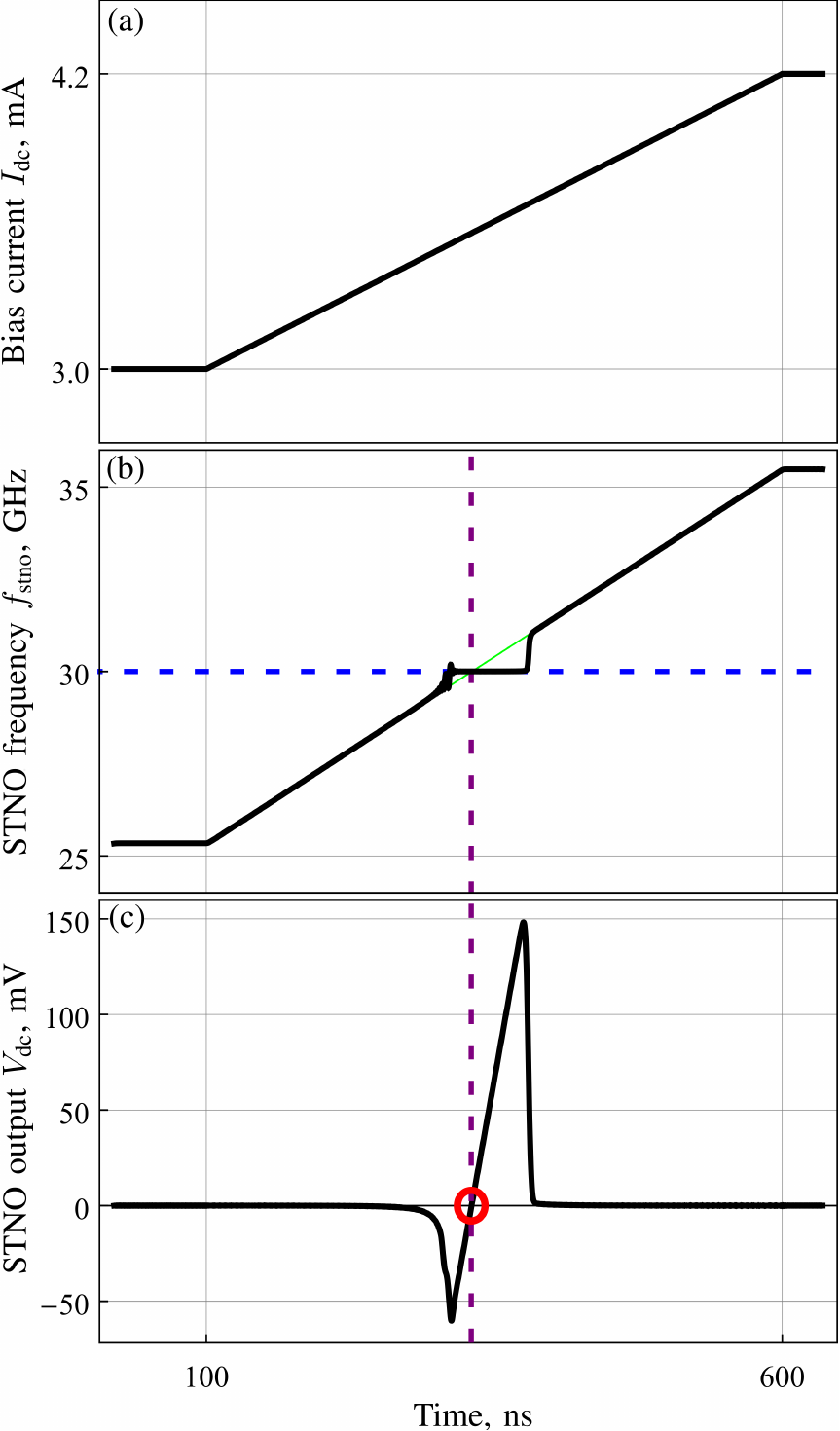}
\caption{Basic operation of STNO spectrum analyzer. (a) The time profile of the ramped bias current, with a 500~ns rise time. (b) The thick black line shows the instantaneous STNO frequency in response to the bias current in (a) and an external microwave signal Eq.~(\ref{irfIn}) with $f_{\rm ext} = 30$~GHz and $I_{\rm rf}=0.2$~mA. Note the injection locking to the external signal, and the otherwise linear increase of the STNO frequency. The free-running STNO frequency in the injection locking interval is shown with a green line. Dashed horizontal line shows the external signal frequency. Dashed vertical line indicates the moment of exact resonance $f_{\rm stno} = f_{\rm ext}$. (c) The output DC voltage of the STNO. In the interval where the STNO is injection locked to the external signal, the STNO produces a sawtooth shaped pulse. Note the pulse crosses the 0~V line at the point of exact resonance.}
\label{basic}
\end{figure}

The basic operation of the STNO spectrum analyzer is demonstrated in Fig.~\ref{basic}. The ramped bias current $I_{\rm dc}(t)$ increased from 3.0~mA to 4.2~mA over 500 ns (Fig.~\ref{basic}(a)). Fig.~\ref{basic}(b) shows the STNO frequency $f_{\rm stno}(t)$ in response to the ramped bias current and input microwave signal with the frequency $f_{\rm ext}=30$~GHz and amplitude $I_{\rm rf}=0.2$~mA. The STNO generates at $\approx 25$~GHz until the bias ramp begins at 100~ns. Then, $f_{\rm stno}$ rises linearly with the scanning rate $R=df_{\rm stno}/dt=0.02$~GHz/ns until it nears $f_{\rm ext}$, where the STNO injection locks to the external signal (see plateau in Fig.~\ref{basic}(b)). As the bias current increases, the SNTO exits the phase locking regime and its frequency resumes linear increase. In the absence of the input microwave signal, the STNO frequency linearly increases in the whole range, which determines the relation between temporal position and microwave frequency.

The STNO voltage output $V_{\rm dc}$ is shown in Fig.~\ref{basic}(c). The output voltage is non-zero only inside the locking interval and has a characteristic sawtooth shape. This specific form of the output DC peak is connected with the variation of the phase shift $\psi$ between the STNO oscillations and the external signal (see Eq.~(\ref{vdc})). The phase shift $\psi$ linearly increases with the STNO free-running frequency \cite{slavin2009nonlinear} and at exact resonance is equal to $\psi = \psi_0 \approx \pi/2$, the intrinsic phase shift of the STNO \cite{zhou2007intrinsic}, which is due to the strong nonlinearity of the STNO. The output voltage reaches maximum value $V_{\rm peak}$ at the right end of the synchronization interval, where $\psi \approx \pi$. Note, that, due to the large intrinsic phase shift of the STNO $\psi_0 \approx \pi/2$, the frequency of the external signal can be precisely determined by digital signal processing (DSP) from zero crossing of the output voltage $V_{\rm dc}(t)$.

\begin{figure}[t]
\centering
\includegraphics{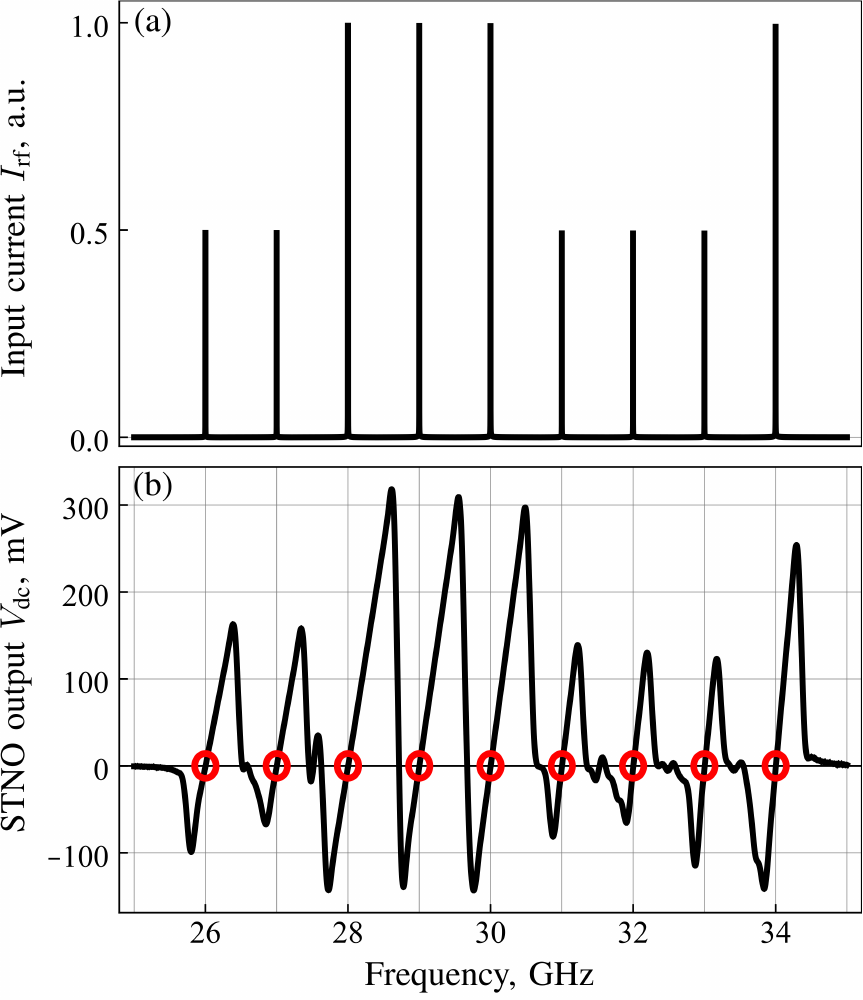}
\caption{Example of spectrum analysis of a complex input signal. (a) Spectrum of input signal consisting of several monochromatic peaks with frequencies between 25 and 35~GHz. (b) Output DC voltage of the STNO spectrum analyzer. Note that the height of each sawtooth pulse is proportional to the amplitude of the corresponding input peak, while the zero crossings, labeled with red dots, coincide with high precision to the input frequencies.}
\label{spec}
\end{figure}

If the STNO is modulated by multiple microwave signals, for example a signal with a spectrum as shown in Fig.~\ref{spec}(a), the STNO will produce spikes of rectified voltage at corresponding frequencies as shown in Fig.~\ref{spec}(b). In Fig.~\ref{spec}(a), the external signal has frequencies at integral values between 25 and 35~GHz. Figure~\ref{spec}(b) shows output STNO voltage $V_{\rm dc}(t)$ mapped to the frequency domain $f_{\rm stno}$. One can see that the STNO faithfully reproduces the complex input spectrum -- the peak voltages $V_{\rm peak}$ are proportional to the amplitudes of the corresponding frequency components of the input signal, while the zero crossing of each sawtooth (indicated by red dots in Fig.~\ref{spec}(b)) precisely matches each input frequency. There is a slight change in the relative amplitudes of $V_{\rm peak}$ related to the change of the oscillating STNO resistance $\Delta R_{\rm stno}$ with bias current. This change, however, is rather weak and can be easily compensated by DSP.


\begin{figure}[t]
\centering
\includegraphics{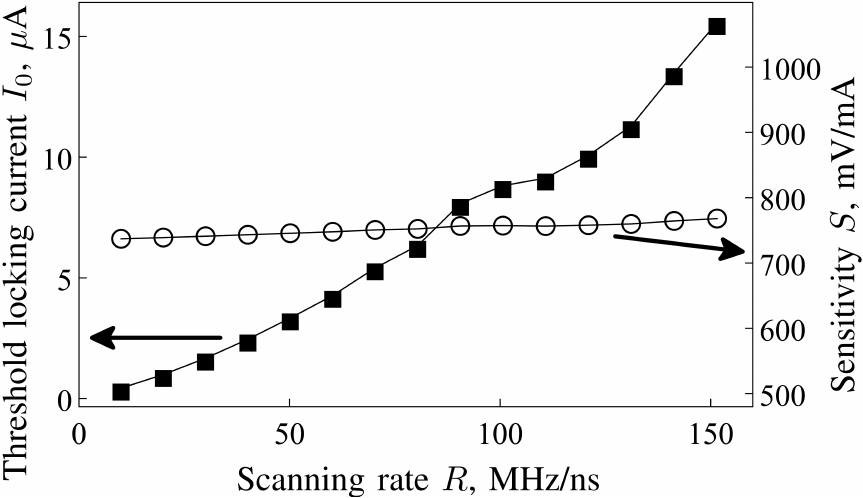}
\caption{Dependence of main characteristics of STNO spectrum analyzer on the scanning rate $R$. Solid squares (left axis): the threshold phase-locking current $I_0$. Open circles (right axis): the STNO sensitivity $S$.}
\label{ithsen}
\end{figure}

Our simulations have shown that at small values of $I_{\rm rf}$, noticeable distortions of the output sawtooth-like STNO peak appear. In this regime the generated DC pulse becomes dependent on the initial phase of STNO oscillations and the peak voltage $V_{\rm peak}$ reduces. A reliable detection of microwave signals is impossible if $I_{\rm rf} < I_0$, where $I_0 = I_0(R)$ is an apparent phase-locking threshold, which strongly depends on the scanning rate $R$. In the region $I_{\rm rf} > I_0$ the peak voltage $V_{\rm peak}$ is accurately described by the simple relation
\begin{equation}
	V_{\rm peak}=S(I_{\rm rf}-I_0)
\label{vpeak}
\,,\end{equation}
where $S = dV_{\rm peak}/dI_{\rm rf}$ is the sensitivity of the STNO detector. The physical origin of the phase-locking threshold $I_0$ is clear: establishing phase-locking between an STNO and an external signal requires certain time $\tau_{\rm pl}$, which is inversely proportional to the signal amplitude $I_{\rm rf}$ \cite{zhou2010oscillatory, rippard2013time} and, if the STNO frequency is scanned over the locking interval faster than $\tau_{\rm pl}$, phase-locking becomes impossible. It is interesting to note that the influence of frequency ramp on injection locking of an oscillator, described by Eq.~(\ref{vpeak}), is analogous to the influence of thermal noise, where apparent locking threshold has been observed experimentally \cite{demidov2014synchronization}.

Fig.~\ref{ithsen} shows the dependence of the sensitivity $S$ and the threshold current $I_0$ on the scanning rate $R$ for signal frequency of 30~GHz. The sensitivity remains practically constant, $S \approx 750$~mV/mA, in a wide range of scanning rates. In contrast, the threshold current $I_0$ increases approximately linearly with $R$ and has typical value $I_0 \approx 10$~$\mu$A at $R = 100$~MHz/ns. This increase of $I_0$ is the main factor limiting the practical scanning rate of the STNO spectrum analyzer.

\begin{figure}[t]
\centering
\includegraphics{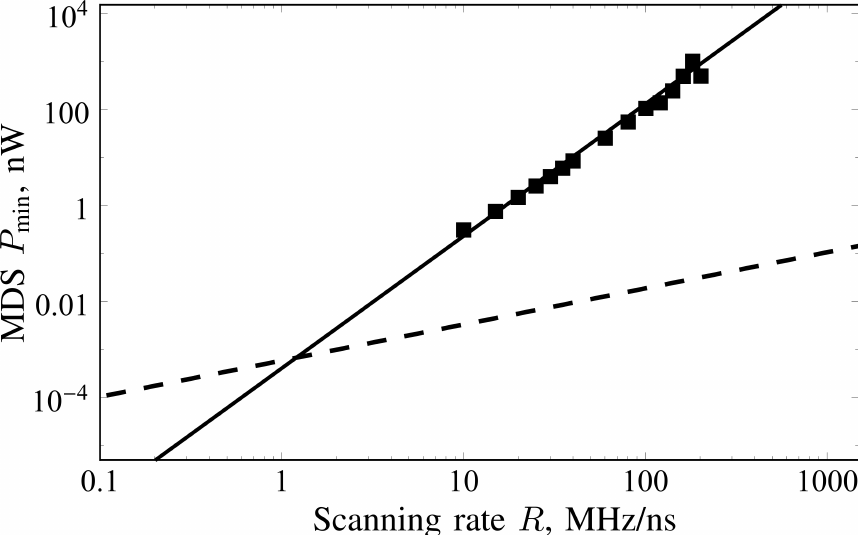}
\caption{Minimum detectable signal (MDS) of STNO spectrum analyzer as a function of the scanning rate $R$. Solid squares: simulated MDS points. Lines: fit of simulated data (solid line) and the thermal noise floor (dashed line). The lines intersect at STNO scanning rate $R=1.2$~MHz/ns and MDS$=0.67$~pW.}
\label{mds}
\end{figure}

The minimum detectable signal (MDS) $P_{\rm min}$ of the STNO spectrum analyzer can be estimated as the input signal power, for which the output voltage Eq.~(\ref{vpeak}) becomes equal to the thermal Johnson-Nyquist (JN) voltage in the bandwidth of low-pass filter $\Delta f$. The dependence of $P_{\rm min}$ on the scanning rate $R$ for $f_{\rm ext} = 30$~GHz is shown in Fig.~\ref{mds} by solid squares. In the range of simulated scanning rates, the MDS is dominated by the threshold current $I_0$ and can be estimated simply as $P_{\rm min} \approx R_0I_0^2/2$. The influence of JN noise (see dashed line in Fig.~\ref{mds}), however, becomes more important with the reduction of the scanning rate and at $R \approx 1$~MHz/ns the two contributions becomes approximately equal. At these rates the MDS is about 1~pW and, thus, the STNO spectrum analyzer can be used as an ultra-sensitive microwave signal detector. In this theoretical work, we have assumed perfect impedance matching. However, in an implemented experiment, good impedance matching over a 10 GHz bandwidth is difficult, and will cause an increased MDS at unmatched frequencies.

\section{Conclusion}

A novel type of ultrafast spectrum analyzer is proposed and investigated theoretically through numerical simulation. The analyzer is based on injection locking of an STNO driven by a ramped bias current. The spectrum analyzer faithfully reproduced spectra of complex incident signals and can have an operational bandwidth of 10~GHz and frequency scanning rate exceeding 100~MHz/ns. The minimum detectable power of the analyzer decreases with the decrease of the scanning rate at is about 1~pW at a scanning rate of 1~MHz/ns.

\section*{Acknowledgment}
This work was supported in part by the Grant No. EFMA-1641989 from the National Science Foundation of the USA, by the contract from the US Army TARDEC, RDECOM, and by the grant from the Center for NanoFerroic Devices (CNFD) and Nanoelectronics Research Initiative (NRI).

\ifCLASSOPTIONcaptionsoff
  \newpage
\fi



\bibliographystyle{IEEEtran}
\bibliography{STNORampBib}
\end{document}